\documentclass[aps,pre,twocolumn,superscriptaddress,citesort]{revtex4-1} %,preprint
\usepackage{color}
\usepackage{graphicx}
\usepackage{CJK} %chinese characters
\usepackage{amsmath}
\usepackage{epstopdf}
\usepackage[normalem]{ulem}
\usepackage{bm}

\begin{document}

	\begin{CJK*}{UTF8}{} % Use default fonts from CJK (see below)
		%\preprint{}
		
		\title{
			Rotational dynamics of bottom-heavy rods in turbulence\\ from experiments and numerical simulations\\
			%Rotational dynamics of non-homogeneous rods in turbulence: experiments and numerical simulations
		}
		%\sout{differentially heated} 
		% address or url should go in the {}'s for \email and \homepage.c
		

		\author{Linfeng Jiang}
		\affiliation{Center for Combustion Energy, Key Laboratory for Thermal Science and Power Engineering of Ministry of Education, Department of Energy and Power Engineering, Tsinghua University, Beijing, China}
		
		\author{Cheng Wang}
		\affiliation{Center for Combustion Energy, Key Laboratory for Thermal Science and Power Engineering of Ministry of Education, Department of Energy and Power Engineering, Tsinghua University, Beijing, China}
		
		\author{Shuang Liu}
		\affiliation{Center for Combustion Energy, Key Laboratory for Thermal Science and Power Engineering of Ministry of Education, Department of Energy and Power Engineering, Tsinghua University, Beijing, China}
		
		\author{Chao Sun}
		%\CJKfamily{gbsn} %needed for local build
		\email{chaosun@tsinghua.edu.cn}
		\affiliation{Center for Combustion Energy, Key Laboratory for Thermal Science and Power Engineering of Ministry of Education, Department of Energy and Power Engineering, Tsinghua University, Beijing, China}
		\affiliation{Department of Engineering Mechanics, School of Aerospace Engineering, Tsinghua University, Beijing 100084, China}
		
		\author{Enrico Calzavarini}
		\email{enrico.calzavarini@univ-lille.fr}
		\affiliation{Universit\'e de Lille, ULR 7512 Unit\'e de M\'ecanique de Lille - Joseph Boussinesq (UML), F 59000 Lille, France}

		%\thanks{}

		\date{\today}
		
		\begin{abstract}
			We successfully perform the three-dimensional tracking in a turbulent fluid flow of small asymmetrical  particles that are neutrally-buoyant and bottom-heavy, \textit{i.e.}, they have a non-homogeneous mass distribution along their symmetry axis. We experimentally show how a tiny mass inhomogeneity can affect the particle orientation along the preferred vertical direction and modify its tumbling rate. The experiment is complemented by a series of simulations based on realistic Navier-Stokes turbulence and on a point-like particle model that is capable to explore the full range of parameter space characterized by the gravitational torque stability number and by the particle aspect ratio. We propose a theoretical perturbative prediction valid in the high bottom-heaviness regime that agrees well with the observed preferential orientation and tumbling rate of the particles. We also show that the heavy-tail shape of the probability distribution function of the tumbling rate is weakly affected by the bottom-heaviness of the particles.
	\end{abstract}
	%\pacs{}
	\maketitle
\end{CJK*}
%\section{Introduction}

\textit{Introduction}
Many turbulent natural and artificial fluid flows are seeded with small inclusions either organic or mineral or manufactured, called dispersed phase. As examples one can mention the water droplets, the ice crystals, the sand grains or the volcanic ashes carried by atmospheric winds; the wide variety of planktonic organisms transported by the oceans \cite{GuastoARFM2012}; and in the industry the liquids loaded in cellulose fibers  in paper making production \cite{2011papermaking} or the tiny  bubbles rising in column chemical reactors. Although the common aspect of all these examples is that they concern the fluid transport of small-in-size inclusions (from now on dubbed particles), each case differs by the specific nature of the particles and by their physical properties. The particles shape can be regular like a sphere or irregular, their material structure can be rigid or deformable, the mass density can be lighter/heavier than the surrounding fluid, and homogeneous or not. For what concerns living particles they can react to external stimuli (like temperature, light or local acceleration and deformations) through motility or change in orientation or shape.

In the last two decades a vast amount of work has been devoted to the investigation of the dynamics of particles in turbulent flows \cite{la2001nature,voth2002jfm,calzavarini2009acceleration,qureshi2007turbulent,Bala2010ARFM,calzavarini2012impact}. This renewed interest into a classical fluid dynamics problem has notably taken a fundamental point of view, considering idealized turbulent flows with their known universal features and simplified models for the particles, often assumed tiny and spherical. Major advances have been performed thanks to the strategy of combining experiments, numerical simulations and theoretical predictions built on models of developed turbulence  \cite{toschibodenshatzARFM}. 
More recently the studies have attacked the problems of less idealized, real so to say, particles, such as non spherical ones \cite{VothARFM2017} or active particles \cite{ardeshiri2017copepods,Calzavarini2018microprobes,Cencini2019Review}.
Researches on the rotational motion of spherical and non-spherical particles \cite{PumirWilkinson2011,ParsaPRL2012,ParsaPRL2014,Gustavsson2014,ni_kramel_ouellette_voth_2015,Byron2015,CandelierPRL2016,KramelPRL2016,GustavssonPRL2017,Bakhuis2019,calzavarini2020anisotropic,jiang_calzavarini_sun_2020,Memarzadeh2018,Molaei2019,KramelPRL2016} have revealed how the complex dynamics of the turbulent velocity gradient in the vicinity of the particle affects the characteristic tumbling and spinning rates. On the other the rotation of larger particles have been connected to the property of turbulence in the so-called inertial range \cite{Shin2005,ParsaPRL2014,BounouaPRL2018}.
Studies of particle dynamics in wall-bounded flow have provided new insights on the fact that particles tumbling properties can be affected by the background shear-flow in a non trivial way
\cite{jiang_calzavarini_sun_2020,Zhao2015,CuiJFM2020}.

When particles are non-homogeneous in their mass density distribution, an extra torque induced by the gravitational field comes into play and it can affect the preferred orientation and rotation of the particle. This has been revealed to be particularly important in the biological domain for tiny motile plankton, leading to the phenomenon called gyro-taxis or gravi-taxis. This phenomenon first proposed in \cite{Kessler1985Nature} is responsible for plankton clustering at small scale by turbulence, and has been the subject of recent important works  \cite{Durham2009scicence,Durham2013NC}.
However, the experimental studies of non-homogeneous active particles by means of living organism are delicate because of their sensitivity to diverse environmental factors and for the unavoidable variability present in any real population. It is clear that experiments with real microorganisms are not the ideal setting to test dynamical models for e.g. non-homogeneous non-spherical particles.

In this paper, we study the statistical properties of the orientation and rotation of passive bottom-heavy neutrally buoyant rod-like particles 
in developed turbulence by means of dedicated experiments and simulations. On one hand this is a challenging experimental task, because of the many control parameters involved and the high accuracy required for the measurements. On the other hand the availability of simplified model system allows for a numerical study of the system that can be checked against experiments. 
We will show that this combined approach is capable to assess the predictive potential of currently adopted models for the Lagrangian dynamics of single bottom-heavy anisotropic particles and to reveal the main trends as a function of the control parameters.

%\section{Setups and methods}
%\subsection{Numerical simulations}
\begin{figure}[!ht]
	\begin{center}
		\includegraphics[width=1\columnwidth]{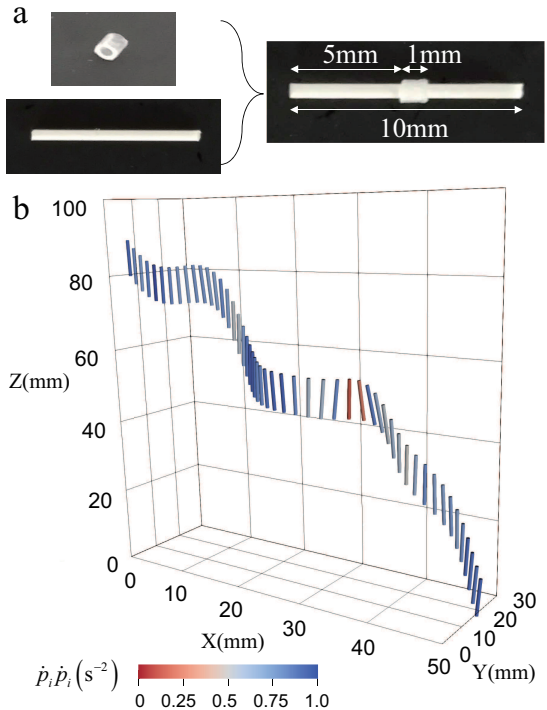}
		\vspace{-0.5cm}
		\caption{a) Photos and scales in mm of one of the axisymmetric bottom-heavy rods and its assembly parts used in the experiments. b) Example of trajectory of a bottom-heavy rod reconstructed via 3D-PTV. The temporal duration of this trajectory is $23 ~sec$. The color encodes the quadratic tumbling rate. 
		}
		\label{fig:trajectory}			
	\end{center}
\end{figure}
%\subsection{Experiments}

\textit{Experimental methods} We carry out the experiments in the central region (referred as bulk) of a Rayleigh-B\'enard convection  (RBC) cell in turbulent flow conditions. The RBC system we consider is cubic with side H=24 cm, while the bulk is chosen as a cube at its center of size $H/3$. The latter choice, already adopted in our previous work \cite{jiang_calzavarini_sun_2020} is motivated by the fact that we have verified that in that region the small scale properties of the turbulent flows share a strong similarity with the ones found in homogeneous and isotropic flows.
A detailed description of the experimental setup can be found in Ref.~\cite{jiang_calzavarini_sun_2020}. Here, we carefully match the density of the working liquid (a solution of 15 \% in weight glycerol in water at a mean temperature $T_m=40^oC$) to the one of the tracked particles so that the particles results to be on average neutrally buoyant in the fluid. The experiments are conducted at Rayleigh number $Ra=\beta g \Delta T H^3/(\nu\kappa)=1.4\times10^{10}$ and Prandtl number $Pr=\nu/\kappa=13$ (here $\beta$ is the thermal expansion coefficient, $g$ is the gravity acceleration intensity, $\Delta T$ the bottom-top temperature difference, $\kappa$ the thermal diffusivity, $\nu$ the fluid viscosity). The global energy dissipation rate in the system is evaluated from the relation $\epsilon=RaPr^{-2}(Nu-1)\nu^3/H^4$\cite{ShraimanSiggia1990}, while the local energy dissipation in the bulk is estimated by considering the ratio of the local energy dissipation rate to global energy dissipation rate via matching with numerical simulations as done in Ref.~\cite{jiang_calzavarini_sun_2020}. Thus, the dissipative length and time scales in bulk are, respectively, $\eta=0.94$ mm and $\tau_\eta=0.87$ s, and the estimated Taylor-Reynolds number in the cell bulk is $Re_{\lambda} \approx 37$.

The bottom-heavy particles are fabricated by threading a polyamide rod into a thin polyethylene ring. The diameter and length of rod are $d=0.56$ mm and $l=10$ mm respectively, which results in an aspect ratio $\alpha=l/d \simeq 18$. The ring has an inner diameter of 0.56 mm and outer diameter 0.96 mm  and a length of 1 mm. The centroid of the ring is placed at the position 0.5 mm away from the centroid of the rod, see particle picture in Figure \ref{fig:trajectory} b). Given the basic geometry of assembled particle, the position of its baricenter can be easily estimated. We estimate  as $h = 0.14$ mm the off-center displacement of the particle centroid with respect to the particle geometrical center.

The particles are tracked using a three-dimensional particle tracking velocimetry (3D-PTV) technique by means of two orthogonally positioned cameras. The details of the tracking method are the same as in Ref.~\cite{jiang_calzavarini_sun_2020}.
We track simultaneously $\sim 20$ particles. Given the volume of the convective cell they can be considered as highly diluted and non-interacting. The recorded series can be as long as a minute, which is in the order of the large eddy turnover time of the flow. A reconstructed trajectory for a typical case is shown in Figure \ref{fig:trajectory} b).

The rotation of particles is governed by the dimensionless stability number $\psi=B/\tau_\eta$ where $B$ is a reorientation time scale due to gravity, which is defined as $B=\nu\alpha_\perp/(2hg)$. Here, 
%$h = 0.14$ mm the off center displacement of the particle centroid with respect to the geometrical center, 
$\alpha_\perp$ the dimensionless resistance coefficient for rotation, same as in \cite{Pedley1990jfm}, which is  $\simeq 207$ in our experiments. 
We note that due to slight differences in the particles, a precise evaluation of $h$ and so of $B$ is arduous. For this reason we also evaluate $B$ in an alternative way, by the direct measurement of the tumbling trajectory in a quiescent flow. In such case the exact evolution of the particle vertical component $p_z$ can be obtained analytically (by solving the equation of motion ~(\ref{eq:Jeffery3d})):
\begin{equation}
p_z(t)  =   \frac{p_z(0)(e^{t/B}+1)+e^{t/B}-1}{p_z(0)(e^{t/B}-1)+e^{t/B}+1}, \label{solution}
\end{equation}
where $p_z(0)$ is a predefined initial condition, that can be easily controlled in the experiment. We proceed as follow, the particle is released in a bottom-up position in the still fluid, its trajectory filmed and the parameter $B$ deduced from the matching with (\ref{solution}). With this procedure we estimate $\psi=0.52 \pm 0.1$, in good agreement with the a priori estimation of B. 
\begin{figure}
	\begin{center}
		\includegraphics[width=1\columnwidth]{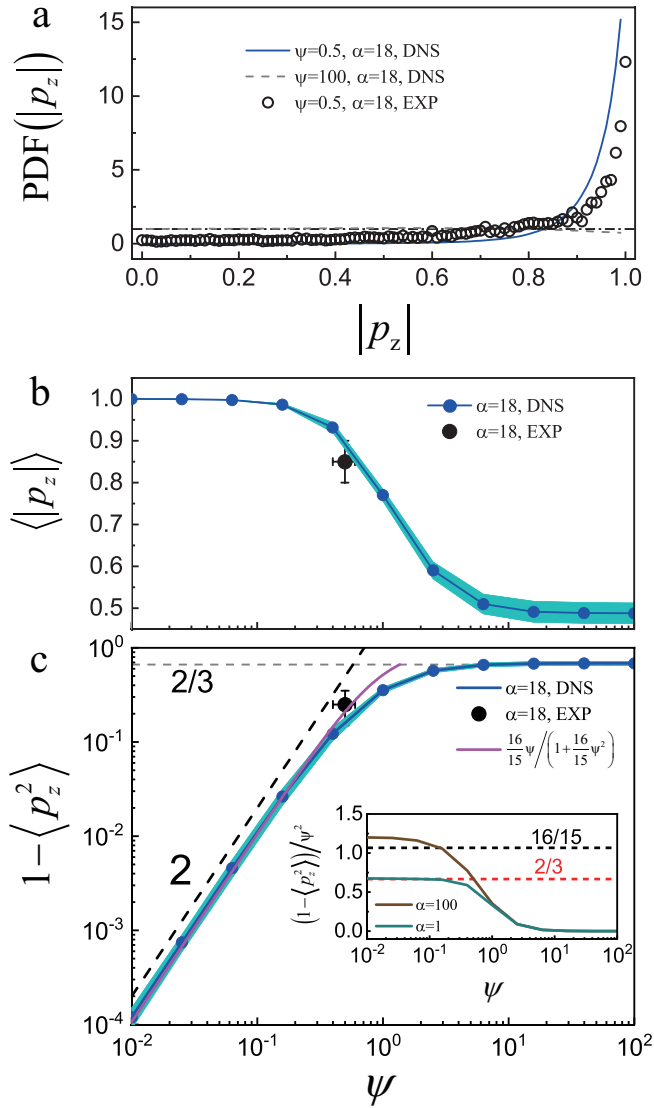}
		\vspace{-0.5cm}
		\caption{
			a) PDFs of absolute value of particle vertical component $|p_z|$ from experiments (EXP) and  simulations (DNS). The black dot-dashed line represents the PDF of randomly oriented particles.
			b) Mean absolute value of particle vertical component $\langle|p_z|\rangle$ as a function of stability number $\psi$ from experiments (EXP) and simulations (DNS). The light blue shaded area represents the error of $\langle|p_z|\rangle$ in simulations. c) $1-\langle p_z^2\rangle$ as a function of $\psi$ in log-log scale. The pink solid line shows the prediction eq. (10). Inset shows $1-\langle p_z^2\rangle$ compensated by $\psi^2$ as a function of $\psi$ and the expected values according to the perturbative predictions (\ref{2/3eq}) and (\ref{16/15eq}). 
		}
		\label{fig:Pz}			
	\end{center}
\end{figure}

\textit{Numerical methods} 
We carry out numerical simulations based on an Eulerian-Lagrangian modelization of the physical problem. This means that the turbulent environment is described by means of direct numerical simulations (DNS) of incompressible Navier-Stokes equations while the particles are described by a 
point-particle model that takes into account the movement of the center of mass and the spatial orientation of the particle. 
The equation for the fluids reads as follows:
\begin{eqnarray}
\partial_t\textbf{u}  + \textbf{u}\cdot {\nabla}   \textbf{u} &=& -   \rho^{-1}{\nabla} p + \nu\ \nabla^2   \textbf{u} + \textbf{f}, \label{eq:N-S}\\
{\nabla}  \cdot  \textbf{u} &=& 0, \label{eq:div1}
\end{eqnarray}
where $\textbf{u}(\textbf{x},t)$ denotes the fluid velocity vector field, $p$ is the hydrodynamic pressure, and parameters are the kinematic viscosity $\nu$, the reference liquid density $\rho$. The vector $\textbf{f}$ refers to an external large-scale random force with constant global energy input which produces and sustains a statistically homogeneous and isotropic turbulent (HIT) flow. Periodic boundary conditions are enforced along all the directions of the three-dimensional cubic simulation domain. The HIT flow can be characterized by a single dimensionless control parameter the Taylor-Reynolds number $Re_{\lambda}=u_{rms}\lambda/\nu$ where $u_{rms}$ is the single-component root-mean-square velocity and $\lambda=\sqrt{15\nu/\epsilon}u_{rms}$ is the Taylor micro-scale of turbulence. We simulate an HIT flow at $Re_\lambda \approx 32$ on a regular grid of $128^3$ nodes. 

The model for the Lagrangian evolution of a single particle position, $\textbf{r}(t)$, and orientation, $\textbf{p}(t)$, is described by the following equations:
\begin{eqnarray}
\dot{\textbf{r}} &=& \textbf{u}(\textbf{r}(t),t),\label{eq:part}\\
\dot{\textbf{p}} &=&\frac{1}{2B}[\textbf{z}-(\textbf{z} \cdot \textbf{p})\textbf{p}]+\Omega \textbf{p} +\tfrac{\alpha^2-1}{\alpha^2+1} \left( \mathcal{S}\textbf{p} - (\textbf{p}\cdot \mathcal{S}\textbf{p})  \textbf{p} \right),\quad \label{eq:Jeffery3d}
\end{eqnarray}
where $\textbf{z}$ is a unit vector pointing upward (i.e. opposite to the gravity direction), 
$\mathcal{S} = ({\nabla}   \textbf{u}+ {\nabla}   \textbf{u}^T)/2$ and $ \Omega=({\nabla}   \textbf{u}-{\nabla}   \textbf{u}^T)/2$ respectively represent the symmetric and anti-symmetric components of the fluid velocity gradient tensor at the particle position, ${\nabla}   \textbf{u}$. The equation of rotation (\ref{eq:Jeffery3d}) is an extension of the Jeffery equation\cite{Jeffery1922} which considers the gravity torque due to the center of mass offset in the particle, as proposed by Pedley \& Kessler \cite{Pedley1990jfm}. This model
has been originally proposed in the context of motile algae studies and since then extensively used to study the dynamics of gyrotactic swimmers\cite{Durham2013NC,Cencini2019Review}. This model is appropriate for particles whose motion is ruled by hydrodynamics in the viscous regime. It applies to turbulent flows as long as the particle size and the particle translation and rotation response times are of the order of the respective dissipative (i.e. Kolmogorov) scales. Although here we have $d\simeq 10 \eta$, the response times are $O(\sim0.1)\tau_{\eta}$ which provide still a reasonable condition for the applicability of the model (a fact also confirmed past studies on fibers in turbulence \cite{Bakhuis2019}).
The simulated HIT flow is seeded by a large amount of particles ($\sim 10^6$), divided into distinct families characterized by  aspect ratio and stability number parameters varying in the range $\alpha \in [0.01,100]$ and $\psi \in [0.01,100]$.
The simulations are performed with the {\sc Ch4-project} code \cite{Calzavarini_SI2019}.\\

%\subsection{Results}

\begin{figure}
	\begin{center}
		\includegraphics[width=1\columnwidth]{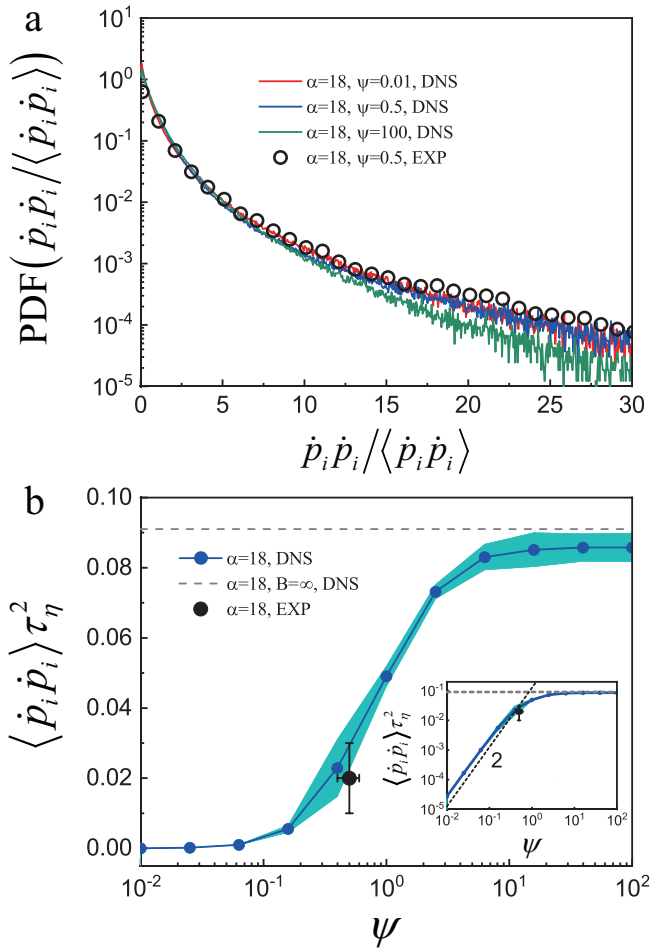}
		\vspace{-0.5cm}
		\caption{a) PDFs of tumbling rate squared $\dot{p}_i\dot{p}_i$ from experiments (EXP) (circles) and simulations (DNS) (full lines). b) Normalized mean tumbling rate squared $\langle\dot{p}_i\dot{p}_i\rangle\tau^2_{\eta}$ as a function of stability number $\psi$ from EXP and DNS. The light blue shaded area represents the statistical error of $\langle\dot{p}_i\dot{p}_i\rangle\tau^2_{\eta}$ in simulations. Grey dashed line refers to the value of mean tumbling rate squared for homogeneous rod with a similar aspect ratio at $Re_\lambda=32$ The inset show the same data on a log-log scale, the $\psi^2$ behaviour is shown as black dashed line. 
		}
		\label{fig:Tumb}			
	\end{center}
\end{figure}
\textit{Results: Orientation} 
We begin discussing the effect of preferential orientation along the vertical direction of the particles, which appears to be a dominant effect in the experimental setting  as it can be noticed in the reconstructed trajectory of Figure \ref{fig:trajectory} b).
This can be quantified by the absolute value of particle z-axis component $|p_z|$. We use the absolute value here in order to allow for the comparison between experimental measurements and simulations, in fact the experimental trajectory reconstruction technique allows to detect the particle direction but not to resolve its orientation, due to the tiny asymmetry of its body. The probability density function (PDF) of $|p_z|$ is expected to be non-uniform in the range $[0,1]$. This is illustrated in Fig.~\ref{fig:Pz} a) which reports the experimental measurement for the rod-like particle with aspect-ratio $\alpha=18$ and stability number $\psi = 0.5$. In the same figure we show the results of simulations for the same parameters, which are in reasonable quantitative agreement with the experiment. For comparison we also provide the case for high stability number $\psi=100$, where the particle orientation is only weakly affected by the gravity, and the distribution appears essentially flat. 
\begin{figure}
	\begin{center}
		\includegraphics[width=0.9\columnwidth]{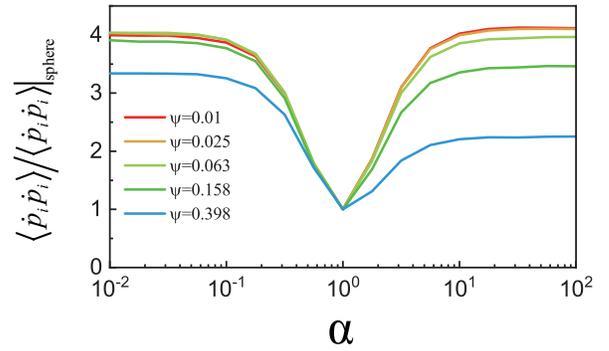}
		\vspace{-0.5cm}
		\caption{The ratio of mean squared tumbling rate of non-spherical particles with respect to the sphere case, $\langle \dot{p}_i\dot{p}_i\rangle/\langle \dot{p}_i\dot{p}_i\rangle_{\alpha=1}$, at various stability number $\psi$ in the limit of highly stabilized particles ($\psi \to 0$).
		}
		\label{fig:Fig5}			
	\end{center}
\end{figure}

The trajectory and ensemble (i.e. particle population) average for the same quantity, $\langle|p_z|\rangle$, as a function of the stability number $\psi$ is shown in Figure~\ref{fig:Pz} b). Such a quantity, as illustrated by the DNS results, monotonically decreases as $\psi$ increases from a gravity torque dominated regime $(\psi\ll1)$ to the turbulence dominated regime $(\psi\gg1)$. The value $\psi\sim 1$ identifies an intermediate state between the two limiting behaviours (here $\langle|p_z|\rangle \sim 1/\sqrt{2})$ meaning $45^o$ angle with respect to the vertical), confirming that the dissipative time-scale of turbulence $\tau_{\eta}$ is the appropriate scale for the description of the dynamics of a tiny particle in turbulence. It also appears that a plateau is reached in the two limits when $\psi \lesssim 0.1$ or $\psi \gtrsim 10$.
The experimental measurement agrees with the simulations and show a strong alignment with the vertical; if we convert the observable to an average angle in degrees   acos$(\langle|p_z|\rangle)  \simeq 32^o$.

In the limit of small but non-vanishing $B$, a perturbative solution of equation (\ref{eq:Jeffery3d}) up to the first order in $B$  can be performed. We consider the Taylor expansion of the solution $\mathbf{p}$ around the point $B=0$: $\mathbf{p}(t) \simeq \mathbf{p}_0(t) + B \mathbf{p}_1(t) + \mathcal{O}(B^2)$, and substitute it into (\ref{eq:Jeffery3d}). The equation is then solved order by order in $B$ starting by the leading order. This leads to 
\begin{equation}\label{perturb-small}
\mathbf{p} \simeq  \mathbf{z} + B (\mathbf{\bm \omega} \times \mathbf{z} + 2 \tfrac{\alpha^2 - 1}{\alpha^2 + 1}   \left( \mathcal{S}  \mathbf{z} - \mathcal{S}_{zz} \mathbf{z} \right) ) + \mathcal{O}(B^2).
\end{equation}
We note that this prediction does not satisfy the constraint of unit norm for $\mathbf{p}$, hence its normalized expression reads:
\begin{eqnarray}\label{perturb-notsmall}
\mathbf{p} &\simeq& \frac{\mathbf{z} + B (\mathbf{\bm \omega} \times \mathbf{z} + 2 \tfrac{\alpha^2 - 1}{\alpha^2 + 1}   \left( \mathcal{S}  \mathbf{z} - \mathcal{S}_{zz} \mathbf{z} \right) ) }{\sqrt{1 + B^2 (\mathbf{\bm \omega} \times \mathbf{z} + 2 \tfrac{\alpha^2 - 1}{\alpha^2 + 1}   \left( \mathcal{S}  \mathbf{z} - \mathcal{S}_{zz} \mathbf{z} \right) )^2 }} 
\end{eqnarray}
However, eq. (\ref{perturb-small}) can still be considered as an acceptable approximation of the solution in the very small $B$ limit, because it is the first order Taylor expansion of eq. (\ref{perturb-notsmall}).
These derivations when combined with the statistical properties of the velocity gradient tensor in HIT flows $\langle \partial_a u_b \partial_c u_d \rangle \tau_{\eta}^2= (4\delta_{ac}\delta_{bd} - \delta_{ab}\delta{cd} - \delta_{ad}\delta_{bc} )/30$\cite{hin75},
allow to estimate the mean quadratic orientation along the vertical direction $\langle p_z^2 \rangle$. We discuss here two special cases: the sphere $\alpha =1 $ and the thin rod $\alpha \to \infty$.
In the first case ($\alpha =1 $), from (\ref{perturb-small}), we get 
\begin{equation}\label{2/3eq}
1-\langle p_z^2\rangle \simeq  B^2(\langle\omega_y^2\rangle+\langle\omega_x^2\rangle)=\frac{2}{3}\psi^2,
\end{equation}
in the second case ($\alpha \to \infty$), again from (\ref{perturb-small}), 
\begin{equation}\label{16/15eq}
1-\langle p_z^2\rangle \simeq  B^2(\langle\ (\partial_z u_x)^2\rangle+\langle(\partial_z u_y)^2\rangle)=\frac{16}{15}\psi^2.
\end{equation}
We observe that from eq. (\ref{perturb-notsmall}), upon averaging and by means of Jensen inequality one can derive the slightly more general relation
\begin{equation}\label{16/15uneq}
1 - \langle p_z^2\rangle \leq \frac{\frac{16}{15}\psi^2}{1+\frac{16}{15}\psi^2}.
\end{equation}
Such predictions are compared with the measurements in Figure~\ref{fig:Pz} c), where it is clearly seen the quadratic trend for $1 - \langle p_z^2\rangle \sim \psi^2$ for $\psi <1$ and the saturation $1 - \langle p_z^2\rangle = 2/3$ occurring in the regime dominated by turbulent fluctuations  $\psi \gg 1$. The predictions for thin rods (\ref{16/15eq}) and (\ref{16/15uneq}) compare well both with the DNS and experimental results. The agreement on the scaling prefactor, which is excellent in the case of the sphere, is compared in the inset of  Figure~\ref{fig:Pz} c).

\textit{Results: Tumbling} We now focus on the tumbling rate of the particle, i.e. on its rotation rate in the direction orthogonal to the symmetry axis. Previous studies have shown that the particle shape is responsible for a  phenomenon of preferential alignment of rod-like particles with the fluid vorticity leading for prolate particles to a reduced tumbling rate with respect to the spherical case and the opposite for oblate particles \cite{Shin2005, PumirWilkinson2011,ParsaPRL2012}. We show here how the gravitational torque further reduces the tumbling of the particles.

The PDFs of tumbling rate squared for different $\psi$ are shown in Figure~\ref{fig:Tumb} a). The long tail of PDF of tumbling rate has also been observed for homogeneous axi-symmetric particles\cite{ParsaPRL2012,ParsaPRL2014,jiang_calzavarini_sun_2020}. This denotes the presence of intense local rotation rates with respect to the average value $\langle\dot{p}_i\dot{p}_i\rangle$. The PDF from experiments agrees with the PDF from simulation up to about $\sim 10$ standard deviations. The differences observed for higher tumbling rate regime could be a signature of sub-leading inertial effects associated to the finite-size of the particles, as observed in \cite{ParsaPRL2014} for homogeneous fibers. Furthermore, a weak stability number dependence is observed on the tail of PDFs, with an increasing intermittency for low $\psi$; such an observation can be considered as preliminary and will deserve further checks at different Reynolds numbers.
\begin{figure}
	\begin{center}
		\includegraphics[width=1\columnwidth]{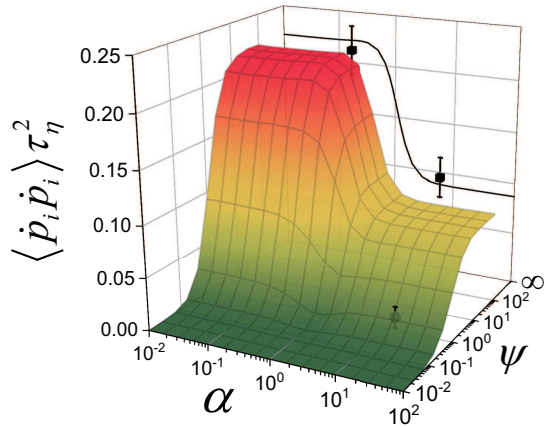}
		\vspace{-0.5cm}
		\caption{Normalized mean tumbling rate squared $\langle\dot{p}_i\dot{p}_i\rangle\tau^2_{\eta}$ as a function of stability number $\psi$ and aspect ratio in EXP and DNS. The color represents the value of $\langle\dot{p}_i\dot{p}_i\rangle\tau^2_{\eta}$. The black circle denotes the data from the present  experiments. The black square  shows the result for homogeneous particles ($\psi=\infty$) of aspect ratio $\alpha=6$ and  1/6 in the RBC bulk at a comparable turbulent intensity, from \cite{jiang_calzavarini_sun_2020}. 
			%For conciseness, it is shown at the position $(\alpha=6$ $ \&$ 1/6, $\psi=100)$.
		}
		\label{fig:Tumb_surface}			
	\end{center}
\end{figure}

The quadratic tumbling rate normalized by the temporal dissipative scale, 
$\langle \dot{p}_i \dot{p}_i \rangle \tau_{\eta}^2$ for $\alpha =18$ particles at varying the stability number $\psi$ is reported in 
Figure~\ref{fig:Tumb} b). We observe a marked suppression of the tumbling rate for the particle in the experiments to about $20 \%$ of the known value for a corresponding equal-in-shape and homogeneous particle. The numerics suggest that the trend of tumbling as a function of $\psi$ is similar to the one already observed for the orientation, with apparent saturation for $\psi$ outside the range $[0.1,10]$.
The perturbative solution (\ref{perturb-small}) when derived with respect to time and averaged gives, for spheres ($\alpha =1$):
\begin{equation}\label{eq:tumb1}
\langle \dot{\mathbf{p}}  \dot{\mathbf{p}} \rangle|_{sphere} \simeq 4 B^2\left(\langle \left( \tfrac{D}{Dt} \partial_i u_j \right)^2 \rangle - \langle  \tfrac{D}{Dt} \partial_i u_j  \tfrac{D}{Dt} \partial_j u_i  \rangle \right)
\end{equation}
and for rods ($\alpha \to \infty$)
\begin{equation}\label{eq:tumb2}
\langle \dot{\mathbf{p}}  \dot{\mathbf{p}} \rangle|_{rod} \simeq 8 B^2 \langle \left( \tfrac{D}{Dt}\partial_i u_j \right)^2 \rangle,
\end{equation}
where the summation over repeated indexes is not implied.
We remark that in both cases the scaling $\langle \dot{p}_i \dot{p}_i\rangle \tau_{\eta}^2 \sim\psi^2$ is expected and well verified in the inset of Figure~\ref{fig:Tumb} b), where we compare it with the numerically measured behaviour for a particle with $\alpha=18$. The average values in eq. (\ref{eq:tumb1}) and (\ref{eq:tumb2}) can not be computed on the basis of purely statistical symmetry arguments. However, they have been numerically measured in \cite{Fang_PhysFluids2015} as $ \langle  \tfrac{D}{Dt} \partial_i u_j  \tfrac{D}{Dt} \partial_j u_i  \rangle \approx 0.6 \langle \left( \tfrac{D}{Dt}\partial_i u_j \right)^2 \rangle$ suggesting a larger tumbling rate for rods as compared to spheres for vanishing $\psi$ values.
This is consistent with our numerical results, where we obtain $\langle \dot{\mathbf{p}}  \dot{\mathbf{p}} \rangle_{rod} / \langle \dot{\mathbf{p}}  \dot{\mathbf{p}} \rangle_{sphere} \simeq 4$ (with respect to the 5 implied by DNS in \cite{Fang_PhysFluids2015}). The described trend is reported in Fig. \ref{fig:Fig5}, for the smallest $\psi$ values explored in our simulations. 

Finally, we address the behaviour of the mean quadratic tumbling rate in the full two-dimensional
parameter space $(\alpha,\psi)$. This can be conveniently done by means of the simulations. The results are illustrated in Fig.~\ref{fig:Tumb_surface}.  At large $\psi$ values, the mean tumbling rate shows two plateaus for slender rods and thin disks, which agrees with the rotation dynamics of homogeneous particles in turbulence ($\psi=\infty$) and with our previous experimental measurements   at $\psi=\infty$ and $\alpha=6$ and $1/6$  in the same bulk of RBC setting \cite{jiang_calzavarini_sun_2020}.  As the stability parameter decreases, one observes a more pronounced similarity between disk-like and rod-like particles. 
And, as already remarked in Fig. \ref{fig:Fig5} in that limit the tumbling both of rods and thin disks is larger than the one of corresponding spheres, in agreement with the expectation from the perturbative prediction.\\

\textit{Conclusions} In summary, we presented a study of the statistics of orientation and rotation of anisotropic bottom-heavy particles in a turbulent flow by means of experiments and simulations. The effect of gravity induced torque due to the inhomogeneity of particles, characterized by stability number $\psi$, significantly modifies the statistics of the particle orientation and its rotation intensity. At low $\psi$, the particles are brought to align with the gravity direction. It is found that the in this limit  $1-\langle p_z^2\rangle$ and mean tumbling rate squared $\langle \dot{p}_i\dot{p}_i\rangle$ scale as $\psi^2$ at small $\psi$. A prediction based on perturbation theory shows excellent agreement with the DNS measurements. Our results point to the fact that particulate matter with  geometrical and mass density properties that are different from the idealized homogeneous one, are likely to possess rotational properties that are very different from the one indicated by the homogeneous particle model. As we show here even a tiny mass bottom-heaviness stabilizes the particle and drastically reduce its tumbling. Further studies will be needed in future to better understand the rich rotational dynamics of non-ideal particles in turbulent flows. In particular it would be of great interest to extend the present investigation to the analysis of spinning rates (i.e. the rotations around the symmetry axis), to the examination of inertial effects for larger particles and ultimately to a the exploration of much more particle shape types and mass-density asymmetries.

\begin{acknowledgments}
This work was supported by Natural Science Foundation of China under grant no.~11988102.
\end{acknowledgments}
\nocite{*}
%\bibliography{OffcenterRod}


\end{document}